# Dual angular tunability of 2D infrared notch filters: Analysis, experiments, physics

Y. H. Ko[1], K. J. Lee[1], F. A. Simlan[1], N. Gupta[2], and R. Magnusson[1*]

[1]*Department of Electrical Engineering, University of Texas at Arlington, Arlington, Texas 76019, USA*
[2]*Sensors and Electron Devices Directorate, CCDC Army Research Laboratory, Adelphi, MD 20817, USA*
*[*]magnusson@uta.edu*

**Two-dimensional (2D) resonant gratings enable dual angular tunability by controlling the plane of incidence (POI) under linear polarization. If the POI is set to be perpendicular to the electric field vector (s-polarization or transverse electric (TE) polarization), an excited TE mode provides spectral tuning. The orthogonally propagating TM mode is robust in angle. Conversely, if the POI is set for p-polarization, an excited TM mode provides the tuning. Detailed explanations of the underlying physical processes are set forth by decomposing the 2D lattice into equivalent 1D gratings using second-order effective-medium theory (EMT). This is shown to work extremely well even for a strongly modulated lattice with refractive-index contrast of 3. With proper design and corresponding experiments, a widely tunable notch filter covering longwave infrared bands is demonstrated. Experimentally varying the incidence angle up to 16°, a notch channel in TM polarization tunes across a band exceeding 0.5 μm with the TE channel remaining at a constant wavelength. The interesting appearance of a resonance channel originating in diagonally propagating leaky modes is briefly examined. The analysis and experiments presented will be useful for realizing diverse 2D tunable filters while furnishing methodology for detailed understanding of the attendant near fields and mode structure.**

## 1. INTRODUCTION

The long-wave infrared (LWIR) spectral region spanning ~8 to 12 μm is important for many scientific and industrial applications such as long-range terrestrial imaging, night vision, and remote environmental monitoring [1–3]. Especially, in optical processing of LWIR spectra, tunable notch filters are demanded as key components to reject unwanted signals of certain frequencies while passing the remaining signal or image. Whereas various technologies including thin film optics and liquid crystals [4, 5] potentially realize tunable filters, guided-mode resonant (GMR) periodic lattices are superb candidates for fashioning compact and efficient notch filters with lossless dielectric media [6]. Previously, we demonstrated tunability of one-dimensional (1D) GMR notch filters by employing chirped gratings and by angular rotation [7, 8]. In contrast, the tuning properties of 2D lattices remain largely untreated. Here, we ameliorate that situation and formulate the properties underlying the tunability of 2D structures in detail. An advantage of the 2D case relative to 1D filters is that by simple polarization rotation, an entirely new tunable spectral range becomes available.

## 2. RESONANT MODE STRUCTURE

For clear insight into the physical origin of the angular properties of the 2D resonant lattices in focus here, we analyze them in terms of decomposed 1D resonant structures in relevant diffraction mounting. Figure 1 explains how the resonance spectra of a representative 2D notch filter originate mainly in two orthogonal leaky resonance modes. To support this contention, we compare the spectral response of a 2D notch filter in Fig. 1(a) with the response of the quasi-equivalent 1D structures in Figs. 1(b) and (c). As illustrated in Fig. 1(a), the 2D notch filter consists of a 2D square array of germanium (Ge, $n_{Ge}$=4) blocks forming a grating, a homogenous Ge sublayer and a zinc selenide (ZnSe, $n_{ZnSe}$=2.4) substrate in air ($n_a$=1). Ge and ZnSe are commonly used as low loss materials in the LWIR region [9]. As decisive geometric factors relevant to resonant modes, we use grating parameter set {Λ, F, $D_g$, $D_h$} denoting the period (Λ), fill factor (F), grating depth ($D_g$) and sublayer thickness ($D_h$). The color map in Fig. 1(a) presents the calculated zero-order transmittance ($T_0$) spectra of a 2D notch filter {Λ=3.7 μm, F=0.5, $D_g$=1.3 μm}. By varying $D_h$, the $T_0$(λ, $D_h$) is simulated under normal incidence with linear polarization (i.e., electric field E=$E_y$, magnetic field H=$H_x$). For accurate calculation, we perform a rigorous coupled-wave analysis (RCWA) [10]. As seen in $T_0$(λ, $D_h$) of Fig. 1(a), there exist three different notch channels ($T_0$=0), which are caused by guided-mode resonance (GMR) excited by different evanescent diffraction orders. In the main channels, distinct resonances are generated by the (0, 1) and (1, 0) diffracted waves coupled to leaky waveguide modes. Herein, the labelling of (*m*, *n*) means the *m*th and *n*th order of diffraction along the ±x and ±y directions, respectively [11]. When $D_h$ increases, each channel moves to longer wavelengths because the GMR redshifts by increased sublayer thickness [12]. At thick $D_h$>0.7 μm, a third sharp resonance is also excited by a GMR mode with the (1, 1) diffracted wave. Focusing now on the two main channels, we verify the orthogonal leaky resonance with quasi-equivalent 1D gratings by homogenizing the 2D grating to 1D arrays with effective-medium theory (EMT) using the approach presented originally in [13]. In Fig. 1(b), the 2D notch filter is converted to a 1D grating equivalent for (0, 1) diffraction. As appropriate for input polarization of $E_y$, the grating bar is homogenized along the x-axis by second order effective refractive index ($n_{eff}^{TE}$) for TE polarization [14] using

$$n_{eff}^{TE}(\lambda) = \sqrt{n_{TE}^2 + \frac{\Lambda^2}{3\lambda^2}\pi^2 F^2(1-F)^2(n_{Ge}^2 - n_a^2)^2} \quad (1)$$

$$n_{TE} = \sqrt{Fn_{Ge}^2 + (1-F)^2 n_a^2} \quad (2)$$

where $n_{TE}$ is the zeroth-order EMT index for TE polarization. The color map represents the $T_0(\lambda, D_h)$ of the equivalent 1D structure where the grating parameter set is the same as for Fig. 1(a). As compared to Fig. 1(a), the $T_0$ locus of the 1D equivalent resonance system is well matched to the broad resonant channel. We note that the 1D resonant mode is a $TM_0$ mode because the E field is perpendicular to the 1D grating grooves. As well-known for 1D resonant grating structures, the observed $TM_0$ resonance channel is formed by resonant radiation which is mediated by Bloch waves counter-propagating perpendicular to the grating grooves (i.e., ±y direction in this case). Proceeding along similar lines, Fig. 1(c) shows the $TE_0$ GMR spectrum of the equivalent 1D grating where the grating bar is homogenized by $n_{eff}^{TM}$ along the y axis according to

$$n_{eff}^{TM}(\lambda) = \sqrt{n_{TM}^2 + \frac{\Lambda^2}{3\lambda^2}\pi^2 F^2(1-F)^2(n_{Ge}^{-2} - n_a^{-2})^2 n_{TM}^6 n_{TE}^2} \quad (3)$$

$$n_{TM} = 1/\sqrt{Fn_{Ge}^{-2} + (1-F)n_a^{-2}} \quad (4)$$

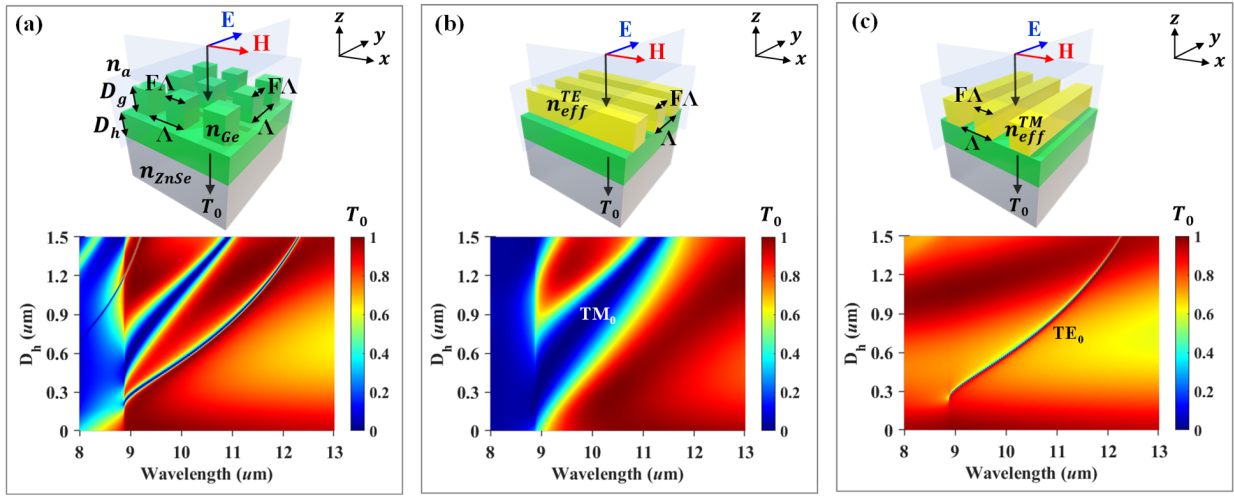

**Fig. 1.** Resonant modes in a 2D notch filter. (a) Ge-based ($n_{Ge}$=4) 2D periodic structure on a ZnSe ($n_{ZnSe}$=2.4) substrate. Grating parameters include period (Λ=3.7 μm), fill factor (F=0.5), grating depth ($D_g$=1.3 μm) and sublayer thickness ($D_h$). At normal incidence under linear polarization (i.e., E=$E_y$, H=$H_x$), the 2D resonant modes are featured in zero-order transmittance ($T_0$) map as a function of the $D_h$. Using effective medium theory (EMT), the 2D modes are decomposed into two quasi-equivalent 1D structures in (b) TM mode (E field perpendicular to grating groove) and (c) TE mode (E field parallel to grating groove). The equivalent 1D structures homogenized by second order EMT ($n_{eff}^{TE}, n_{eff}^{TM}$).

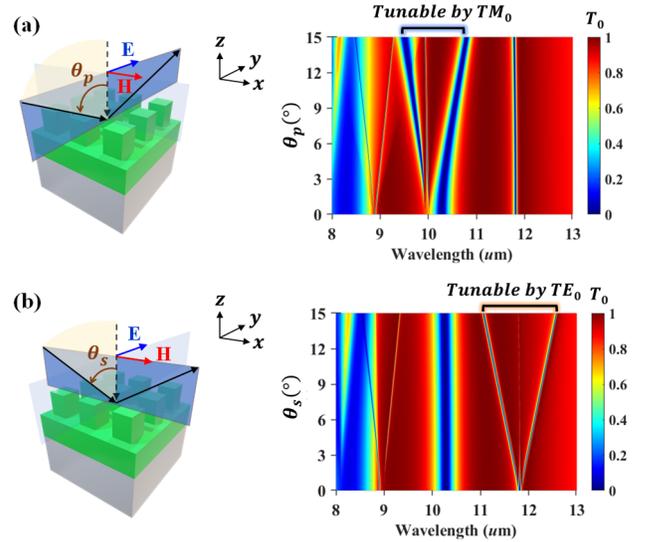

**Fig. 2.** Dual angular tunability of a 2D notch filter. The grating parameter set is {Λ=3.7 μm, F=0.5, $D_g$=1.3 μm, $D_h$=1.21 μm}. The angle of incidence is controlled in two different planes of incidence (POI) under (a) p-polarization ($\theta_p$) and (b) s-polarization ($\theta_s$). In p-/s-polarization, the H-/E-field is perpendicular to the POI. The color maps show $T_0(\lambda, \theta_p)$ and $T_0(\lambda, \theta_s)$. Varying the input angles $\theta_p$ and $\theta_s$, the notch channels can be tunable by way of $TM_0$ and $TE_0$ resonant modes.

where $n_{TM}$ is the zeroth order effective refractive index for TM polarization. In this case, the resonance forms by $TE_0$ modes with Bloch waves propagating along the ±x directions. As seen in the color map, the $T_0$ locus is very similar to the sharp main channel in Fig. 1(a) for the 2D spatial modulation. These results bestow considerable confidence on the EMT-based 1D equivalence methodology to explain spectra of 2D resonant devices.

## 3. ANGULAR TUNING CHARACTERISTICS

We now address how the orthogonal mode propagation just formulated informs the angular properties of 2D resonant filters. As depicted in Figs. 2(a) and 2(b), the input light can be oblique in two different planes of incidence with the first one being transverse to the H field (p-polarization) and the second perpendicular to the E field (s-polarization). The color map of Fig. 2(a) shows the angular spectra of $T_0$ as a function of $\theta_p$ (i. e., angle of incidence under p-polarization) where the grating parameter set is {Λ=3.7 μm, F=0.5, $D_g$=1.3 μm, $D_h$=1.21 μm}. In viewing the $T_0(\lambda, \theta_p)$ map, the $TM_0$ mode implements a tunable channel while the $TE_0$ mode channel remains robust in angle. In fact, at off-normal incidence under p-polarization, the $TM_0$ mode experiences classical mounting while the $TE_0$ mode undergoes fully conical mounting. In general, the classic incidence varies the resonance wavelength strongly with angle because it deviates the diffracted waves much more than in conic incidence as explained in [15]. This is discussed in more detail in Fig. 3. In contrast, in s-polarization as in Fig. 2(b), the $TE_0$ mode supports the tunable channel, now at longer wavelengths, by changing $\theta_s$ (i.e., angle of incidence under s-polarization). Therefore, if both $\theta_p$ and $\theta_s$ are controlled in sequence, the tunable range can be expanded by the orthogonal resonances of the $TM_0$ and $TE_0$ modes; this is a main advantage of applying 2D modulation.

The angular properties of the 2D tunable filters can be further clarified by decomposed resonant modes of quasi-equivalent 1D gratings. As previously treated in Fig. 1 for normal incidence, the 2D grating is homogenized along x and y using $n_{eff}^{TE}$ and $n_{eff}^{TM}$ via EMT with the same grating parameters. Figures 3(a) and 3(b) show the angular properties of the TM and TE mode 1D gratings under p-polarization varying $\theta_p$. Herein, classic/conic incidence refers to the POI as perpendicular/parallel to the grating grooves. In case of the TM mode in classic incidence, as shown in the $T_0(\lambda, \theta_p)$ map of Fig. 3(a), a split $TM_0$ mode correlates well with the tunable channel of Fig. 2(a) that is computed with non-approximated RCWA. On the other hand, in Fig. 3(b), the $TE_0$ resonant mode is tolerant in $\theta_p$ reproducing the vertical channel of Fig. 2(a) at λ=11.8 μm. As discussed in prior works, the differing angular properties of classic and conic incidence can be understood by the variation of the diffracted wave vector $|\Delta k_{diff}| = ||k_{diff}| - |k_{diff}(\theta = 0°)||$, which deviates the GMR wavelength from that at normal incidence [15]. For pertinent explanation, we recall the proper equations of the $|\Delta k_{diff}|$ for the classic ($|\Delta k_{classic}|$) and conic ($|\Delta k_{conic}|$) incidence [15] as

$$|\Delta k_{classic}| = k_0 \sin\theta \tag{5}$$
$$|\Delta k_{conic}| = \sqrt{(k_0\sin\theta)^2 + (mK)^2} - mK \tag{6}$$

where the $m$, $k_0$ and $K$ denote the diffraction order, wave vector ($2\pi/\lambda$) and grating vector ($2\pi/\Lambda$). Using a Taylor series expansion, in the subwavelength regime (Λ < λ), Eq. (6) is approximated as

$$|\Delta k_{conic}| = \frac{2m\pi}{\Lambda}\left\{\sqrt{\left(\frac{\Lambda}{m\lambda}\sin\theta\right)^2 + 1} - 1\right\}$$
$$\approx \sin\theta\left(\frac{\Lambda}{2m\lambda}\right)|\Delta k_{classic}|. \tag{7}$$

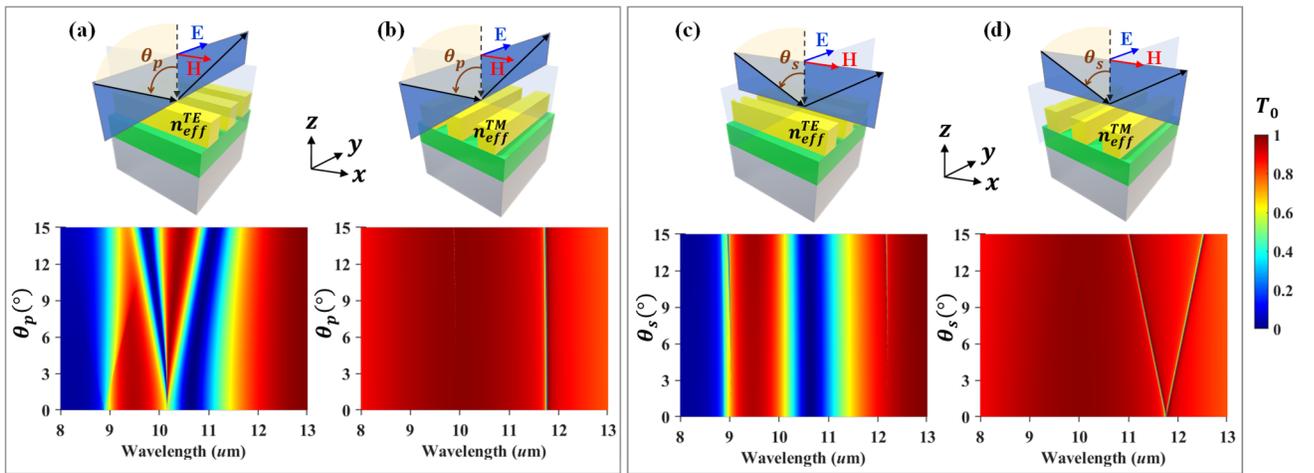

**Fig. 3.** Decomposition of the 2D resonant mode. Angular properties of two quasi-equivalent 1D gratings are characterized under (a, b) p-polarization and (c, d) s-polarization. As for the p-polarization, which is connected to Fig. 2(a), the $T_0(\lambda, \theta_p)$ maps are represented for (a) TM mode under classical mounting and (b) TE mode under fully conical mounting. In the classical/conical mounting, the POI is perpendicular/parallel to grating grooves. For the s-polarization, connected to Fig. 2(b), the $T_0(\lambda, \theta_s)$ maps are shown for (c) TM mode under conical and (d) TE mode under classical mounting.

From Eq. (7), we see that $\Delta k_{conic}$ is smaller than $\Delta k_{classic}$. Therefore, the GMR position is changed slowly by conical incidence due to a smaller variation of the $\Delta k_{conic}$. Figures 3(c) and 3(d) show similar properties of the conical and classical incidence under s-polarization where the structures are TM and TE mode 1D gratings. In the $T_0(\lambda, \theta_s)$ map of Fig. 3(c), the $TM_0$ mode forms a vertical locus as being angularly tolerant at conical incidence. On the other hand, in Fig. 3(d), the $TE_0$ mode under classical mounting is split with a relatively large spectral variation. These two $T_0(\lambda, \theta_s)$ maps correlate well with the 2D angular properties of s-polarization in Fig. 2(b). In particular, the tuned channels in Fig. 3(d) reproduce the rigorously computed 2D result nearly exactly.

## 4. THE DIAGONAL MODE

There appears a curious spectral feature in Fig. 1(a) denoting a narrow resonance locus near $\lambda$ =9 μm and for $D_h$>0.7 μm. It extends into the low-transmission region where the main resonances have been extinguished by appearance of propagating higher diffraction orders at wavelengths below the Rayleigh wavelength $\lambda_R = n_{ZnSe}\Lambda$=8.88 μm. We identify this locus as being resonantly generated by the diagonal (1,1) modes propagating in the resonant structure composed of the 2D lattice and its sublayer. The period seen by this mode is $\Lambda/\sqrt{2}$ reducing the Rayleigh wavelength accordingly and allowing the feature to survive. Figure 4(a) depicts a quasi-equivalent 1D structure and the reduced period ($\Lambda'$) of the diagonal diffraction mode. The diagonal diffraction of (1, 1) or (-1, -1) will be available in the diagonal 1D grating grooves (1, -1) or (-1, 1). Herein, for homogenization, we use the TM mode EMT ($n_{eff}^{TM}$) as appropriate for TE polarized input as noted in Fig. 4(a). Figure 4(b) shows the $T_0(\lambda, D_h)$ map where the grating parameter set is {$\Lambda' = \Lambda/\sqrt{2}$ =2.616 μm, F=0.5, $D_g$=1.3 μm, $D_h$=1.21 μm}. Due to the shortened Rayleigh wavelength ($\lambda_R' = \lambda_R/\sqrt{2}$=6.278 μm), the $T_0$ resonance signature survives in the 8-12 μm LWIR region. The resonance locus generated by the EMT structure shown in Fig. 4(b) agrees

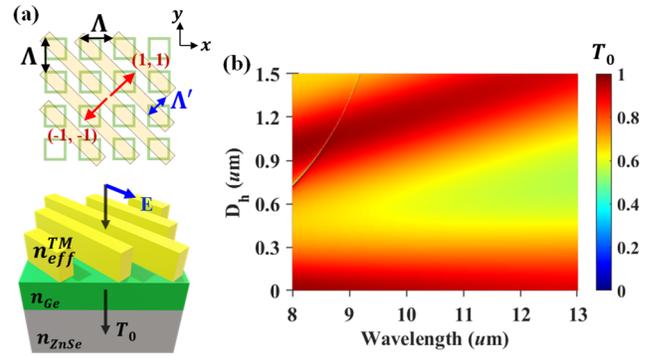

**Fig. 4.** Analysis of the diagonal mode. (a) Homogenized gratings for diagonal diffraction (1, 1) or (-1, -1) where the grating parameter set is {$\Lambda' = \Lambda/\sqrt{2}$ =2.616 μm, F=0.5, $D_g$=1.3 μm, $D_h$=1.21 μm}. For homogenization, we use $n_{eff}^{TM}$. (b) $T_0(\lambda, D_h)$ map of the equivalent 1D structure under TE polarized input at normal incidence.

remarkably well with the diagonal mode feature of the full 2D structure in Fig. 1(a).

## 5. EXPERIMENTAL RESULTS

For the experiments, we fabricated the 2D tunable filters by photolithography and dry etching processes [16]. The ZnSe substrate (with antireflection coated backside for 7-12 μm, Crystran Limited, UK) is prepared and a 2.51-μm-thick Ge film deposited on it via an e-beam evaporator with a deposition rate of 3.0 Å/sec. Then, a spin-coated positive photoresist (PR, AZ MIR 701) is 2D-patterned by UV exposure where we use a 3.7-μm-period and 0.5-fill-factor chromium mask. Thereafter, the Ge film is reactive-ion etched with $CHF_3/SF_6$ gases to 1.3-μm-depth. Finally, the 2D tunable filter is obtained by removing any residual PR. Figure 5(a) displays the device and measurement setup. As depicted in photographic and atomic force microscope (AFM) images, the 2D square array is inscribed as rectangular cells on the 1-inch-substrate. To work with unpolarized as well as with vertically and horizontally polarized input, we prepare the beam path of a Fourier transform infrared spectrometer (FTIR, Nicolet iN10) appropriately. Figure 5(b) shows the measured $T_0$ spectra for the unpolarized and vertically/horizontally polarized input Gaussian beam (~1.5 mm diameter) at normal incidence. Here, the FTIR sample spacing is set to 0.482 cm$^{-1}$ for high spectral resolution. The three measured $T_0$ spectra are almost the same implying a polarization independent 2D structure. The slight difference is caused by an imperfect square array of the fabricated structure, namely slightly different fill factors along x and y. As previously explained, the $TM_0$ and $TE_0$ resonant modes locate at shorter ($\lambda$~10.34 μm) and longer wavelengths ($\lambda$~11.6 μm), respectively. In both polarizations, we see that the $TM_0$ resonance exhibits high rejection efficiency $T_0$< 0.3% and high side bands $T_0$> 90%. For the $TE_0$ mode, meanwhile, the rejection efficiency is somewhat degraded because the narrow bandwidth of the $TE_0$ mode limits the efficiency because of the divergence of the Gaussian beam. Figure 6 shows the measured $T_0$ spectra for different angle of incidence. As illustrated in each diagram, the 2D filter is rotated under illumination of p- and s-polarized input. In viewing Fig. 6(a), as $\theta_p$ increases up to 16°, the $TM_0$ mode splits from $\lambda$= 9.68 μm to 10.97 μm owing to the large variation of diffracted waves in classic incidence. As shown in Fig. 6(b), the simulated tuned $T_0$ spectra match the measurement results reasonably well with main deviations appearing in the efficiency of the sidebands as the theoretical model assumes infinite substrates and therefore no reflection loss at the output interface. In contrast, the $TE_0$ mode does not change because it is at conic incidence. When the angle varies as $\theta_s$, as seen in Fig. 6(c), the $TE_0$ resonance is tunable under classic incidence whereas the $TM_0$ resonance does not tune as it is in conic incidence. At $\theta_s$=16°, the $TE_0$ mode is split from $\lambda$=10.92 μm to 12.46 μm. In Fig. 6(d), the simulated $T_0$ spectra agree with experiment in main features with the greatest deviations seen in the notch efficiency on account of its narrow linewidth under Gaussian beam incidence and attendant divergence. Sharp resonances near $\lambda$~9 μm in the theoretical plots of Figs. 6(a) and 6(b) appear due to the diagonal mode in Fig. 4(b) that exists in this parametric space because the experimental sublayer thickness is near $D_h$=1.2 μm.

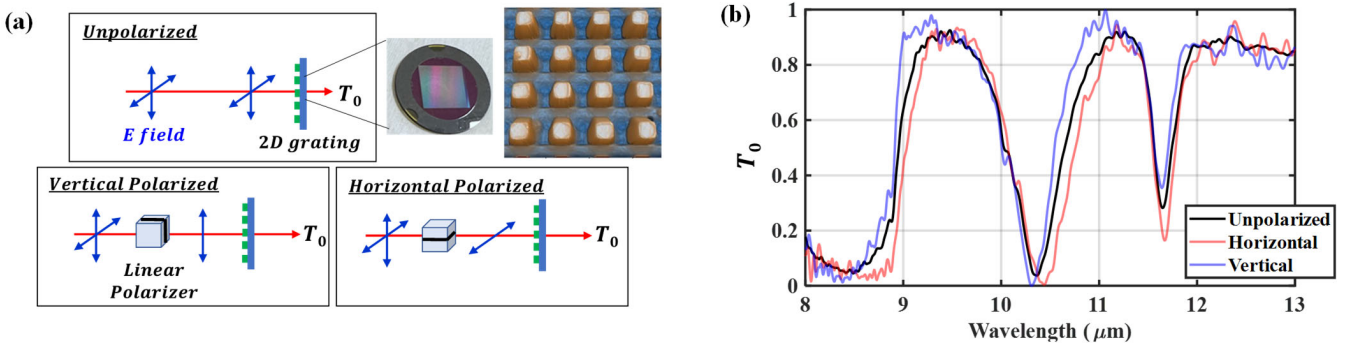

**Fig. 5.** Measurements of the experimental 2D grating device. (a) Using the FTIR with a linear polarizer, the $T_0$ spectra are gathered for three states of polarization, that is unpolarized, vertically polarized, and horizontally polarized input. (b) Measured spectra at normal incidence. The photographic and AFM images of the device appear in (a).

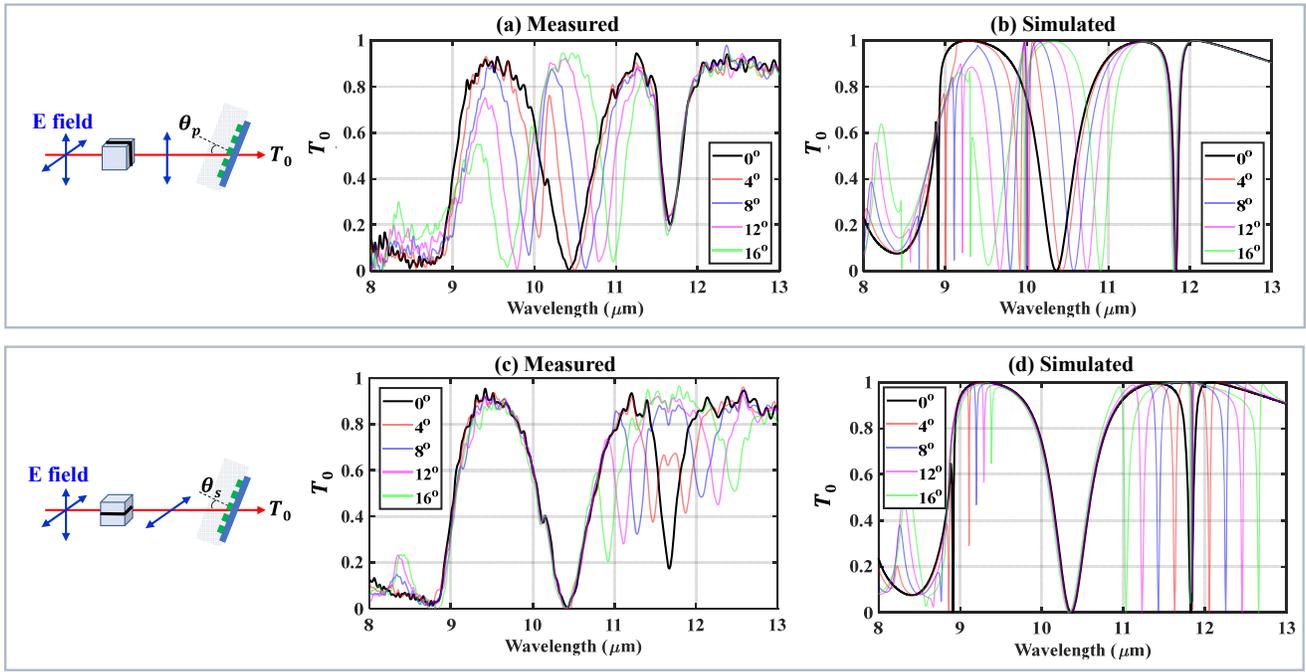

**Fig. 6.** Measurement results for dual angular tunability of the 2D notch filters. The $T_0$ spectra are measured by varying the $\theta_p$ and $\theta_s$ up to 16° in (a) and (c). Two main channels are tunable by $TM_0$ (short wavelength) and $TE_0$ (longer wavelength) resonant modes with the $\theta_p$ and $\theta_s$. Computed results for transmittance versus wavelength are shown for a set of selected angles of incidence in (b) $\theta_p$ and (d) $\theta_s$.

## 6. CONCLUSIONS

The principles and properties of 2D resonant metasurfaces under angular tuning are presented. It is shown that these constructs enable dual angular tunability by choice of the diffraction mounting under linear polarization. Thus, relative to the 1D filters in previous studies, the 2D device provides two tunable spectral ranges that are accessible by simple polarization rotation. The tunable notch filters of interest are implemented by TM and TE resonant modes under effective classical incidence. Meanwhile, under fully conical mounting, the orthogonal resonant mode is tolerant in angle. Accordingly, with a tuned notch channel based on a TM or TE mode, there exists a robust untuned channel grounded in the orthogonally polarized mode. However, these properties are neither explicit nor simple in the case of 2D gratings because the TM and TE resonant modes are blended in spectrum as they undergo mixed diffraction mounting in angular tuning. For resolving them, we apply our effective construction of quasi-equivalent 1D gratings to decompose and reveal the mixed modes as well as the effective diffraction mounting. The simulation and analysis presented here verify the detailed connections between the 2D and the equivalent 1D EMT-based resonances in the tunable 2D notch filters showing the immense utility of this decomposition method. As discovered during this research, this even works well for a diagonal lattice mode that generates a discernable spectral signature in the full 2D resonance system. Furthermore, the experimental results demonstrate and quantify the expandable tunability of Ge/ZnSe-based 2D filters in the LWIR region. The tunability of one mode channel with simultaneous non-tunability of the orthogonal channel is verified experimentally. We remark that by improved design and materials choice, the switched channels can be made more compatible in linewidth than those provided by the example structures presented here; this is desired in most applications. Appreciation of 2D resonances as grounded in the elemental optical properties of 1D resonances affords important insight to understand and develop various types of tunable 2D resonant filters along the lines laid here.

**Funding.** This research was supported by the US Army Research Laboratory under contract W911NF19-2-0171. Additional support was provided by the UT System Texas Nanoelectronics Research Superiority Award funded by the State of Texas Emerging Technology Fund as well as by the Texas Instruments Distinguished University Chair in Nanoelectronics endowment.

**Acknowledgment**. Parts of this research were conducted in the UT Arlington Shimadzu Institute Nanotechnology Research Center.

**Disclosures**. The authors declare no conflicts of interest.

**Data Availability**. Data underlying the results presented in this paper are not publicly available at this time but may be obtained from the authors upon reasonable request.